\documentclass{article}
\usepackage{style/stylo}
\usepackage{graphicx} 

\title{Can Generative AI be Egalitarian?}
\author{Philip Feldman,\footnote{These authors contributed equally.} \ James R. Foulds,$^*$ Shimei Pan}
\date{\today}

\begin{document}

\maketitle

\begin{figure}[h]
    \centering
    \fbox{\includegraphics[width=0.75\linewidth]{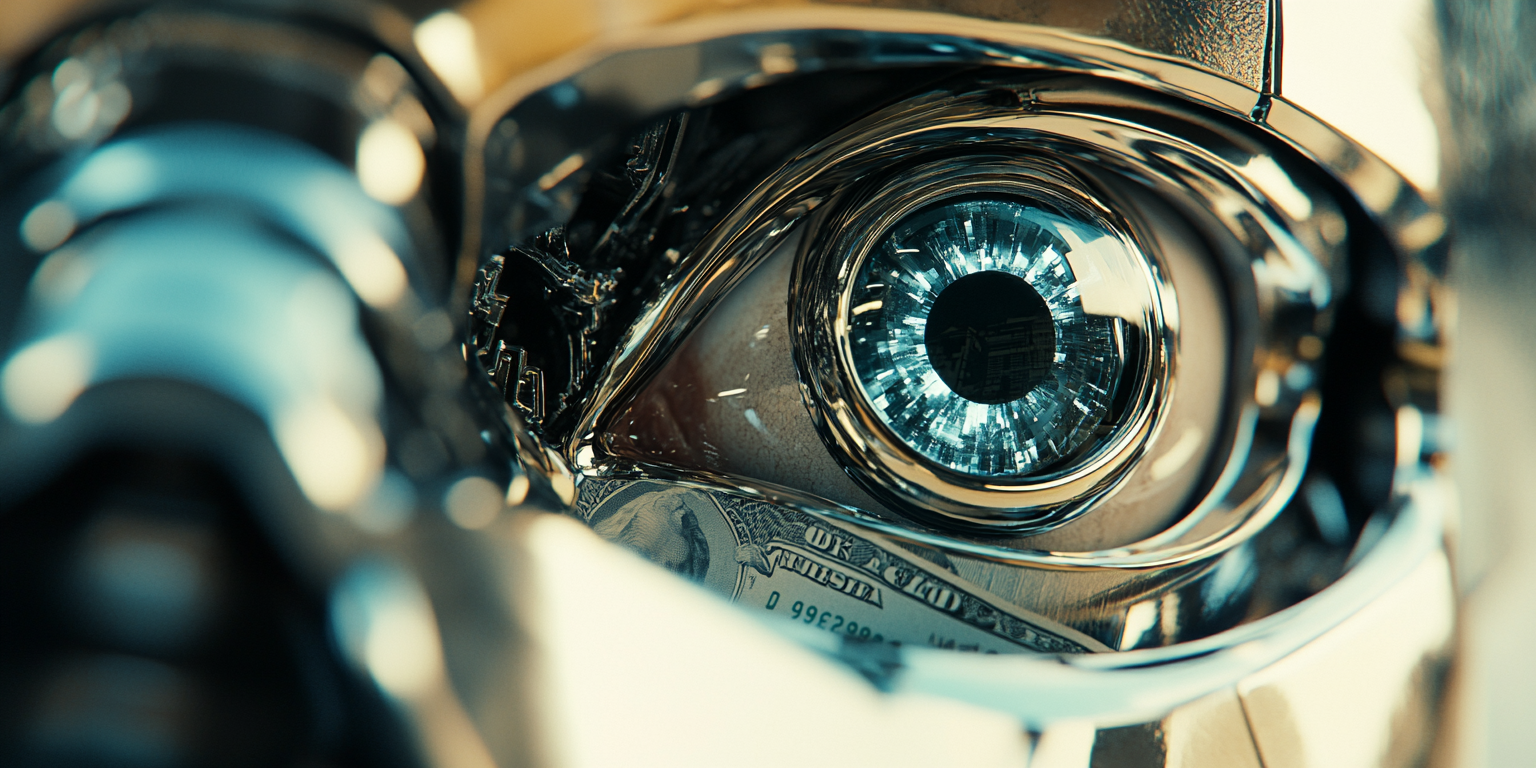}}
    \label{fig:money-robotl}
\end{figure}

\begin{abstract}
The recent explosion of ``foundation'' generative AI models has been built upon the extensive extraction of value from online sources, often without corresponding reciprocation. This pattern mirrors and intensifies the extractive practices of surveillance capitalism~\cite{zuboff2019surveillance}, while the potential for enormous profit has challenged technology organizations' commitments to responsible AI practices, raising significant ethical and societal concerns. However, a promising alternative is emerging: the development of models that rely on content willingly and collaboratively provided by users. This article explores this \enquote{egalitarian} approach to generative AI, taking inspiration from the successful model of Wikipedia. We explore the potential implications of this approach for the design, development, and constraints of future foundation models. We argue that such an approach is not only ethically sound but may also lead to models that are more responsive to user needs, more diverse in their training data, and ultimately more aligned with societal values. Furthermore, we explore potential challenges and limitations of this approach, including issues of scalability, quality control, and potential biases inherent in volunteer-contributed content.

\end{abstract}


\section{Introduction}
Generative AI technologies such as AI chatbots are poised to deeply impact many facets of daily life. These systems are now capable of creating realistic content and answering sophisticated user queries, and are controlled by powerful technology organizations. 

It was not always clear that this would happen, and the trajectory of the AI field to reach this point raises concerns about the future. 
The field of Artificial Intelligence (AI) has experienced a series of booms and busts since its inception in 1956. Early optimism and ambitious predictions led to significant investment and research, but the inability to meet those lofty goals resulted in the first \enquote{AI winter} in the mid-1970s~\cite{mitchell2021ai}. A resurgence of interest in the 1980s, fueled by projects like Japan's Fifth Generation Computer Project~\cite{feigenbaum1993japanese}, led to further advancements but again fell short of expectations, leading to a second AI winter around 1987. Alongside advances in machine learning such as support vector machines and statistical learning, the early 1990s saw the early stages of a resurgence of neural network technologies. Later rebranded as ``deep learning,'' advances in GPU-based training, improvements in models and algorithms, and especially, larger datasets, eventually led to their dominant status from the mid-late 2010s to today.~\cite{chen2016evolution}. These breakthroughs in what has become known as \enquote{generative AI} has developed into a third wave of investment and research.

Throughout these cycles, the funding landscape for AI has also evolved, shifting from reliance on bureaucratic decision-makers in Washington, D.C., and other government entities to the venture capitalists and startups of Silicon Valley, Beijing, London, and Tel Aviv, among others. This shift is not merely geographic; it represents a fundamental change in how machine intelligence is conceived, developed, and deployed.

The US Department of Defense (DoD)'s interests have historically been tied to fulfilling national security needs. Whether the goal was self-driving military vehicles or automated threat detection systems, the emphasis was on trust and reliability. However, the venture capital model, fueled by the desire for rapid and profitable exits, operates with a different ethos. This is particularly evident in Silicon Valley, where a form of arbitrage has become commonplace. In this model, technologies developed with federal subsidies, often originating from defense research, are used to disrupt regulated industries, generating substantial returns for investors~\cite{fan2022nontraditional}.

The current wave of AI development marks a further evolution in this dynamic. The goals are no longer solely technological, but increasingly social and commercial. The focus is on creating systems that can be monetized on a massive scale, with the global population serving as both resource and customer. This shift is facilitated by the vast computational infrastructure that has grown alongside the internet, and the enormous troves of data that can be extracted from it, often without regard for copyright or consent.

\begin{figure}[h!]
    \begin{minipage}{0.33\textwidth}
        \centering
        \fbox{\includegraphics[height=15em]{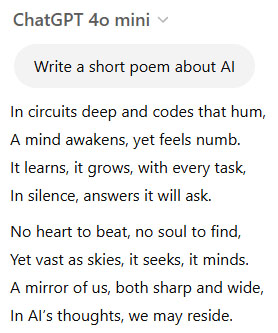}}
        \caption{Chatbot}
        \label{fig:chatgpt}
    \end{minipage}\hfill
    \begin{minipage}{0.33\textwidth}
    \centering
        \fbox{\includegraphics[height=15em]{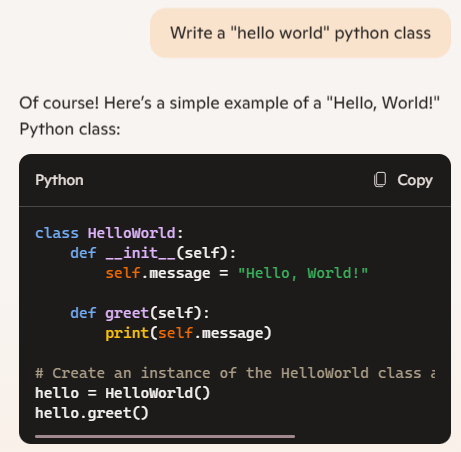}}
        \caption{Coding}
        \label{fig:copilot}
    \end{minipage}
    \begin{minipage}{0.33\textwidth}
    \centering
        \fbox{\includegraphics[height=15em]{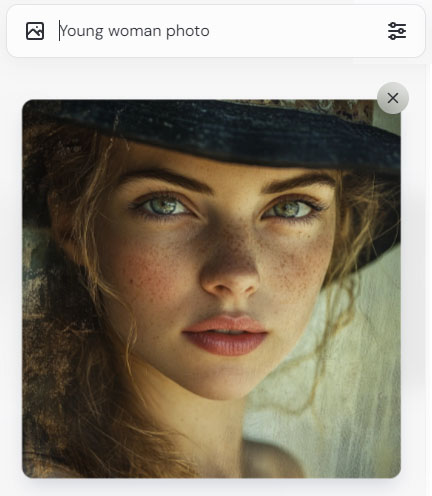}}
        \caption{Image Generation}
        \label{fig:midjourney}
    \end{minipage}
\end{figure}

\begin{figure}[h!]
    \centering
    \fbox{\includegraphics[height=20em]{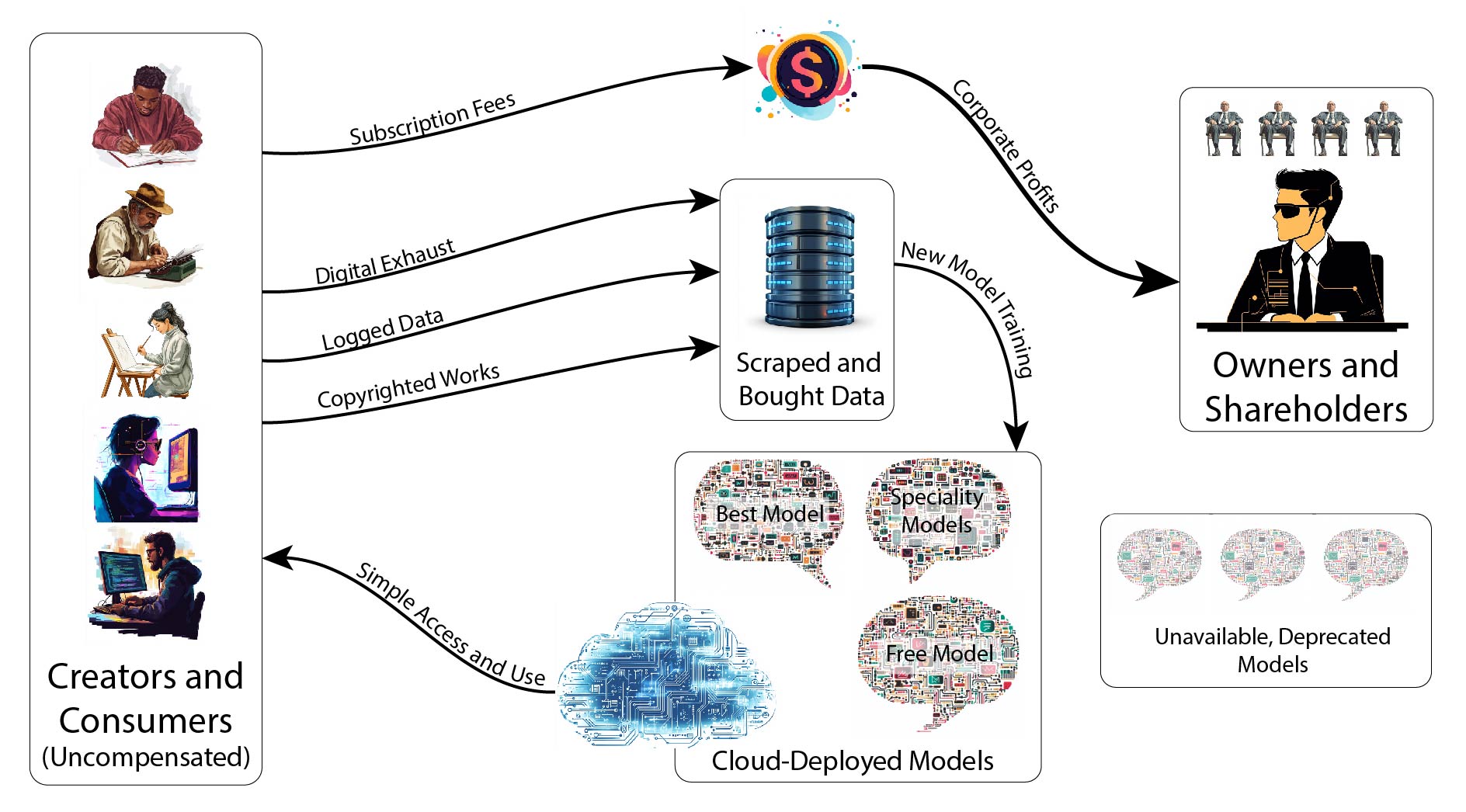}}
    \caption{For-Profit AI Ecosystem}
    \label{fig:for-profit-AI}
\end{figure}

Figure~\ref{fig:for-profit-AI} shows an overview of the for-profit environment as it currently exists in the mid-2020's. In this ecosystem, content creators, often unknowingly, provide their work free of charge. This data is either directly scraped or acquired through intermediaries, as exemplified by Microsoft's 2024 purchase of the Taylor and Francis scientific catalog.\footnote{\url{https://www.thebookseller.com/news/academic-authors-shocked-after-taylor--francis-sells-access-to-their-research-to-microsoft-ai}}  These vast datasets fuel the training of diverse AI models, including chatbots such as ChatGPT and Gemini (Figure~\ref{fig:chatgpt}), coding assistants such as  Copilot (Figure~\ref{fig:copilot}), and image generators such as DALL-E and Midjourney (Figure~\ref{fig:midjourney}).\footnote{All examples generated on 29 December, 2024.} These proprietary models are often inaccessible to the public and are delivered to individual and corporate consumers through cloud-based interfaces, who pay for metered access, typically on a per-token basis.  As models evolve, older versions are decommissioned and may become unavailable to consumers. Revenue generated from consumers and investors flows back to the owners, either as profit or reinvestment to enhance AI capabilities.

The implications of this transition are profound. Machine intelligence is moving away from domains where trust is essential, towards systems where inaccuracies and \enquote{hallucinations} are commonplace~\cite{feldman2023trapping}. This is not merely a technical challenge; it represents a profound change in the relationship between humans and machines. The remainder of this paper will explore these implications and propose potential solutions to the challenges they present. The shift from serving national security needs to exploiting entire populations for profit raises urgent questions about ethics, accountability, and the future of AI. 
We contend that generative AI organizations' profit motivations can and do conflict with the urgent need for these systems to behave in a manner beneficial to society. We provide case studies on OpenAI and Google to highlight this, and we discuss how the theory of surveillance capitalism warns of its dangers~\cite{zuboff2015big, zuboff2019surveillance, zuboff2019age}. 
Addressing these issues is crucial if we are to ensure that machine intelligence serves the broader good, rather than merely a high return on investment for corporations and their shareholders. 
We put forward a possible solution: the Free and Open-Source Software (FOSS) model as an approach to developing generative AI collaboratively and altruistically. We will describe the challenges and limitations around this approach, and discuss how they might be addressed. 

\section{Open AI: A Case Study of a Non-Profit Becoming For-Profit}


The tension between profit motivations and altruistic motivations for generative AI systems has been laid bare in OpenAI's public and ongoing struggle with balancing these goals. OpenAI was founded in December 2015 as a non-profit research company. In its initial press-release, it stated ``\emph{Our goal is to advance digital intelligence in the way that is most likely to benefit humanity as a whole, unconstrained by a need to generate financial return. Since our research is free from financial obligations, we can better focus on a positive human impact}''~\cite{openai.com_1}.

This egalitarian non-profit focus allowed the company to hire a number of top AI researchers at its inception, including renowned AI scientist Ilya Sutskever, who became a co-founder of OpenAI and its chief scientist. However, the company's non-profit status was significantly eroded as early as March 2019, when it announced the formation of OpenAI LP, a for-profit subsidiary operated under the direction of its non-profit parent, now called \enquote{OpenAI Nonprofit}~\cite{openai.com_1}. The for-profit company is legally obligated to pursue OpenAI Nonprofit's mission. OpenAI LP is a \enquote{capped-profit} company, meaning that returns for investors are limited, in this case being capped at 100 times their investment.
This unusual structure aims to strike a balance between profit and egalitarianism, and aims to attract the investment and talent that OpenAI claims are necessary to achieve its goals~\cite{openai.com_1}. It has received criticism, however. An article from the Observer points out that \enquote{OpenAI's profit cap is so high that it might as well not exist}~\cite{observer.com_1}.

OpenAI entered into a strategic partnership with Microsoft shortly after forming its for-profit subsidiary in 2019~\cite{openai.com_1}, and as of March 2024, Microsoft held a 49\% ownership stake and rights to up to 75\% of OpenAI's profits until it recoups its investment in the company~\cite{finance.yahoo.com_1}. The partnership also allows Microsoft to leverage OpenAI's technologies into its products, which it has done with aplomb, e.g., its Azure OpenAI service which serves API access to OpenAI's models using Microsoft's cloud infrastructure~\cite{azure.microsoft.com_1}. In 2024, entrepreneur and former OpenAI board member Elon Musk sued OpenAI and its CEO Sam Altman over breach of contract, alleging that they have abandoned the altruistic, charitable goals that their founding agreement requires in favor of seeking profits, specifically for its biggest investor, Microsoft~\cite{axios.com_1}. 
Microsoft's investment in OpenAI has also brought regulatory scrutiny, with the European Commission, the executive arm of the European Union stating it will be ``looking into'' whether it requires review under the EU Merger Regulation~\cite{ec.europa.eu_1}.  

The brief ouster~\cite{openai.com_1}, and subsequent reinstatement, of OpenAI's CEO Sam Altman in November 2023 exposed fault lines within the company regarding the profit vs. egalitarianism debate~\cite{npr.org_1,nytimes.com_1}. It was reported at the time that OpenAI board members, including Dr. Sutskever, had raised concerns about Altman's push toward rapid commercialization of OpenAI's technologies despite risks and safety concerns~\cite{npr.org_1}. An anonymous letter from former OpenAI employees alleged that \enquote{a significant number of OpenAI employees were pushed out of the company to facilitate its transition to a for-profit model}~\cite{gist.github.com_1}.  Altman's job was saved when 91\% of OpenAI's 770 employees threatened to leave the company unless he was reinstated~\cite{npr.org_1}. Microsoft's CEO, Satya Nadella, applauded the move~\cite{nytimes.com_1}, which apparently indicated that the company approved of Altman and his priorities. Sutskever stepped down from the board after Altman's return, did not return to the office, and left the company in May 2024~\cite{vox.com_1}. Sutskever had co-led OpenAI's ``Superalignment'' team, aiming to ensure that artificial superintelligence technologies developed by the company were aligned with the public benefit, and the new startup company he founded, Safe Superintelligence Inc., has the same goal~\cite{reuters.com_1}. The other co-lead of the Superalignment team, Jan Leike, departed at the same time, which was attributed to growing concerns around OpenAI's commitment to AI safety~\cite{vox.com_1}.

In September 2024, OpenAI's leadership exodus continued, with OpenAI's chief technology officer, Mira Murati, who briefly led the company in Sam Altman's absence, and two other top executives stepping down; insiders associated this with the company's moves toward becoming a for-profit company~\cite{nytimes.com_2}. As of that event, nine of the company's eleven founders had stepped down or taken an extended sabbatical, with only Sam Altman and one other founding executive remaining~\cite{observer.com_2}. One week later, OpenAI completed a fundraising deal that valued the company at \$157 billion USD. According the New York Times, the deal included stipulations that the company restructure itself into a for-profit business within two years, or the funding will be converted to debt~\cite{nytimes.com_3}.   At the time of writing in late 2024, therefore, OpenAI's full transition into a for-profit company appears to be a near certainty. It was reported by the Financial Times that OpenAI is considering restructuring as a public benefit corporation (PBC), a relatively new type of for-profit company which several of its rivals, including Anthropic and xAI, have also adopted, and which would help protect against hostile takeovers by investors. If it became a PBC, OpenAI would be required to balance the interests of shareholders with positive societal impacts, and the interests of other stakeholders including employees and society at large. Regarding the PBC plan, in a guest opinion essay for the New York Times, Andrew Kassoy, co-founder of the egalitarian non-profit network B Lab, wrote ``That’s a good start. But this structure alone does not ensure that OpenAI will be held accountable''~\cite{nytimes.com_6}. To summarize the sequence of events:
\begin{enumerate}
    \item OpenAI's egalitarian mission and non-profit status was useful in the early recruitment of top AI talent;
    \item Non-profit egalitarianism has increasingly been sidelined in favor of for-profit goals;
    \item The company has not been able to reconcile egalitarianism and the pursuit of profit, with most of those who advocated for the former either being forced out or quitting the company as it continues its slide toward a full for-profit status;
    \item Many of the employees who left or were forced out of the company, as well as outside experts, felt that its increasing for-profit direction had compromised its egalitarian goals such as safety and alignment with human values.
\end{enumerate}
This suggests an apparent incompatibility between OpenAI's egalitarian and for-profit goals. Arguably, this bodes poorly for the future of egalitarian ideals in other for-profit generative AI companies as well. In an opinion article in the LA Times,  two MIT professors said that Sam Altman's reinstatement \enquote{\emph{confirms that the future of AI is firmly in the hands of people focused on speed and profits, at the expense of all else. This elite will now impose their vision for technology on the rest of humanity. Most of us will not enjoy the consequences}}~\cite{latimes.com_1}. A Columbia Law School blog post stated that \enquote{Whatever happens in OpenAI’s next chapter, protecting the charitable interests is likely to be a heroic task in the face of the overwhelming profit-making incentives. $\ldots$ If OpenAI fails to follow the requirements of its state nonprofit and federal tax-exempt status, the risk is that [OpenAI's] For-Profit-LP and its for-profit subsidiary would subordinate the nonprofit purpose to their profit-making interests}~\cite{clsbluesky.law.columbia.edu_1}. This is already happening. For instance, in 2024 OpenAI quietly deleted language in its usage policy prohibiting the use of its technology for military purposes, potentially opening up the possibility of contracts with the US Department of Defense or other military organizations or contractors~\cite{theintercept.com_1}.


\section{Google: A Case Study of Responsible AI in a For-Profit} 

The conflict between profit-making and the public benefit has also arisen in other companies. At Google, for example, efforts within the company to ensure ethical and responsible AI principles have led to dramatic clashes with senior management. There is growing awareness that AI systems, especially powerful modern generative AI models, can lead to substantial harm if not designed with care, necessitating responsible AI approaches. Some of the core responsible AI principles include \emph{fairness} (ensuring that AI systems do not encode discriminatory bias or other unjust behavior), \emph{accountability} (ensuring that appropriate human parties are held responsible for negligence, crimes, or other harmful behavior in AI systems), and \emph{transparency} (ensuring that decisions made by AI systems can be understood by impacted parties)~\cite{crawford2019ai}. The challenge of making AI systems behave as desired, called the \emph{AI alignment} problem, is pernicious, especially when the goal is that the system conforms to human values and norms~\cite{zhuang2020consequences}. The goal of profit competes with each of these responsible AI principles, so for-profit companies cannot necessarily be trusted to prioritize them, let alone achieve them. 

Most large technology companies currently employ ethical and responsible AI research teams, and Google is no exception. Issues arise, however, when responsible AI researchers' findings present a challenge to profit-making priorities. At Google, ethical AI researchers Timnit Gebru and Margaret Mitchell collaborated on a research paper in 2020, published in 2021, that critiqued generative AI language models such as those created at their company~\cite{bender2021dangers}. Their work discussed concerns around risks and harms from these models including:
\begin{itemize}
    \item The environmental cost of training these models,
    \item training data that encodes hegemonic worldviews and amplifies societal bias,
    \item static training data that does not keep up with changing societal values,
    \item documentation debt, in which the collection of data outpaces documentation about what that data is,
    \item the willingness of humans to attribute ``intelligence'' to fluent text from systems that are merely replicating patterns in text data without true understanding of it (which they called ``stochastic parrots''),
    \item the resultant automation bias (inappropriate trust in AI systems),
    \item and the potential deliberate misuse of generative AI.
\end{itemize}
A Google manager insisted that the Google researchers remove their names from the paper, or retract it~\cite{nytimes.com_4}. Margaret Mitchell modified her name attribution to ``Shmargaret Shmitchell''~\cite{bender2021dangers}, while Timnit Gebru requested further discussion, and said that she would resign at a later date if Google could not explain its decisions and answer her further concerns. Google responded that it would not do so, and accepted her resignation, effective immediately. Dr. Gebru characterized this decision as a firing~\cite{nytimes.com_4}. Dr. Mitchell was fired by Google around two months later, citing code of conduct and security concerns~\cite{theguardian.com_1}, although the timing might lead one to suspect that the company ``went looking'' for a justification to fire her. 

By definition, for-profit companies must prioritize profit and return to investors over other considerations including responsible AI practices and principles. Speaking to the New York Times about this case, researcher Julien Cornebise said ``\emph{This shows how some large tech companies only support ethics and fairness and other A.I.-for-social-good causes as long as their positive P.R. impact outweighs the extra scrutiny they bring}''~\cite{nytimes.com_4}. 

Later, in 2023, Geoff Hinton, one of the leading pioneers of deep learning, quit Google in order to obtain the freedom to advocate for AI safety. Speaking to the New York Times, Dr. Hinton voiced concerns including misinformation generated by AI, impacts on the job market, autonomous weapons, and perhaps eventually, a threat to humanity due to generative AI gone rogue~\cite{nytimes.com_5}. It is telling that Dr. Hinton felt that he had to leave the company in order to voice these concerns. In 2024, upon winning the Nobel prize in physics for his pioneering work in deep neural networks, Dr. Hinton expressed pride in having a student who fired Sam Altman (i.e., Dr. Ilya Sutskever, who we mentioned above)~\cite{techcrunch.com_1}. 
\section{Generative AI and Surveillance Capitalism}

We have discussed examples of how for-profit motivations can conflict with egalitarian intentions for generative AI companies. With less pure intentions, the results may be far worse. In particular, the theory of \emph{surveillance capitalism}, due to Soshana Zuboff~\cite{zuboff2015big, zuboff2019surveillance, zuboff2019age}, provides a lens for examining problematic aspects of certain Big Tech companies' business models. Surveillance capitalist organizations exploit human experiences and actions as raw data for machine learning models, which they obtain via an extractive, non-reciprocal relationships with those humans. In turn, these companies use their dominance of the technological landscape to nudge, coax, or control people toward profitable outcomes. Zuboff formed her surveillance capitalism theory before the advent of OpenAI's ChatGPT service and other similar AI-based chatbots, but it has proved apt for analyzing this rapidly evolving technological landscape.
 
Generative AI technologies are insatiably data hungry, with the amount of training data currently being the primary limiting factor in their performance, and hence, ability to outperform competitors and attract users. Essentially, increasing the number of parameters (i.e., the model's size and complexity) makes these models more powerful (cf. Fig. 3 in~\cite{chowdhery2023palm}), and larger models require more data in order to train them well. More than ever, this situation steers any profit-driven generative AI company toward a surveillance capitalist business model. To build a generative AI system, the first step is to acquire as much data as possible, at minimal cost, and hence minimal compensation to the data providers, leading to a non-reciprocal extractive data collection process.

Generative AI is fast becoming an essential tool for every facet of daily life, both in the work and home spheres. As such, generative AI companies have users ``over a barrel,'' as it becomes difficult to meaningfully opt out and still be competent and competitive as a worker. It seems almost inevitable that providing personal data will become a condition of using generative AI services. And the better these systems become, the more embedded they are in daily life, the more the generative AI companies will be able to exert control. For example, it will be possible for them to continuously nudge users toward providing access to data on more and more aspects of their lives. Any profit-based generative AI company is motivated to do so in order to out-compete their rivals in AI performance.

Zuboff points out that as Big Tech companies mediate more of our digital and actual lives, they seize the power to monitor and enforce contracts, taking this power from the government, which in a democratic society is chosen by the people~\cite{zuboff2015big}. If a user does not make a car payment on their self-driving car, for example, the manufacturer can instruct the car to drive back to them. As generative AI becomes embedded in more facets of life, the companies that make these technologies will obtain similar powers within each of those facets. For instance, if a user asks an AI chatbot to help design their dream home, it can offer to find contractors to build the home, and facilitate the hiring process, a convenient proposition that many users would likely accept. As the mediator, it is in a position to censure  the contractor for not completing the work under the AI company's own imposed time frame, potentially denying access to future work on the platform. It can also block access to its services, which have now likely become essential to daily life, if the user does not pay on time, a transaction in which the AI company takes a cut. Now, consider a near future in which AI chatbots mediate transactions in this manner in most spheres of life. The AI company has then inserted itself into a position of power and control throughout our entire society. This is the surveillance capitalists dream, and a nightmare for the rest of us.

\section{The Egalitarian Alternative}



The Free and Open-Source Software (FOSS) movement presents an alternative approach to for-profit generative AI. FOSS is grounded in the idea that software should be accessible, modifiable, and distributable with minimal restrictions.  At its core, FOSS champions the freedom to use, study, share, and improve software for any purpose. This philosophy is driven by values of collaboration, transparency, and user empowerment, leading to a sea change in software development. Where once large scale systems were built and sold by for-profit entities, now a mixture of individuals and organizations work together using communal structures such as GitHub to create high-quality software that serves diverse needs.

The \enquote{Free} in FOSS, however, often leads to confusion. In this context, \enquote{Free} refers primarily to freedom, not necessarily to price.  While some FOSS is indeed free of charge (\enquote{Free as in (a complementary) Beer}), the core idea is that users have the freedom to use, modify, and distribute the software like one would have the freedom to share a beer with friends. This freedom extends to examining the \enquote{recipe} (source code) and even brewing your own version (modifying the software) to suit your tastes.

The FOSS movement has produced some of the world's most important software, from the Linux operating system to the Apache Foundation, which provides everything from web servers to databases. These projects are embraced by for-profit entities for their utility and cost-effectiveness~\cite{kamp2024free}. This symbiotic relationship, where the community-driven development of open-source software benefits both individuals and corporations, suggests a potential path for open-source Generative AI models such as Large Language Models (LLMs). As profit margins in the generative AI space narrow due to increased competition and the rising costs of training and maintaining proprietary models, the appeal of open-source alternatives, with their collaborative development and transparent training data, may grow~\cite{goth2024AIopen}.

Such models, trained on content willingly provided by users or organizations, could lead to a more equitable and transparent AI ecosystem. The collaborative creation of knowledge exemplified by Wikipedia, where individuals contribute their expertise to build a free and accessible encyclopedia, provides a potential blueprint~\cite{bruckman2022should}. A large-scale, organic approach to \enquote{Egalitarian AI Foundation Models,} drawing on content explicitly contributed for this purpose, could shift the power dynamics in AI development, empowering individuals and communities to participate in the creation and governance of these powerful tools.

The reliance on volunteered and open-source content might initially limit the scope and diversity of training data, potentially impacting the model's performance on certain tasks or in specific domains. However, it could also encourage the creation of specialized, community-driven models tailored to specific languages, cultures, or areas of expertise, fostering a richer and more diverse AI landscape that caters to a wider range of needs and perspectives. The transparent nature of such models, with their open training data and collaborative development, could also enhance trust and accountability, addressing concerns about bias, misuse, and the concentration of power that plague many proprietary models.

Additionally, as competition intensifies among major AI providers like OpenAI, Google, and Amazon, the pricing strategies for their models have become increasingly aggressive. This \enquote{race to the bottom} may lead to unsustainable business models. Many AI companies have yet to establish profitable revenue streams, relying heavily on investor funding to sustain operations. If the trend continues, it could undermine the financial viability of these companies, leading to a potential collapse similar to past technological bubbles~\cite{wong2024AIbubble}.

As the limitations and ethical concerns of the surveillance capitalism model become increasingly apparent, the allure of collaborative, transparent, and community-driven AI development may prove compelling. The future of generative AI may lie not in the extraction of value from the digital commons, but in its collective cultivation, fostering a more equitable, accessible, and accountable AI ecosystem that benefits society as a whole. 

In this egalitarian model, users would actively participate in the creation and/or curation of training data, ensuring that the models are representative of diverse perspectives and less susceptible to biases inherent in data scraped from the internet. The Wikimedia Foundation, the non-profit organization behind Wikipedia, demonstrates the viability of this approach, relying on a global community of volunteers to create and maintain one of the largest and most comprehensive knowledge repositories in the world. The egalitarian Generative AI model would prioritize transparency and user control, allowing individuals to understand and influence the training process and the underlying algorithms. This approach would foster a sense of ownership and empowerment, ensuring that the benefits of Generative AI are shared equitably among its users.

\begin{figure}[h!]
    \centering
    \fbox{\includegraphics[height=20em]{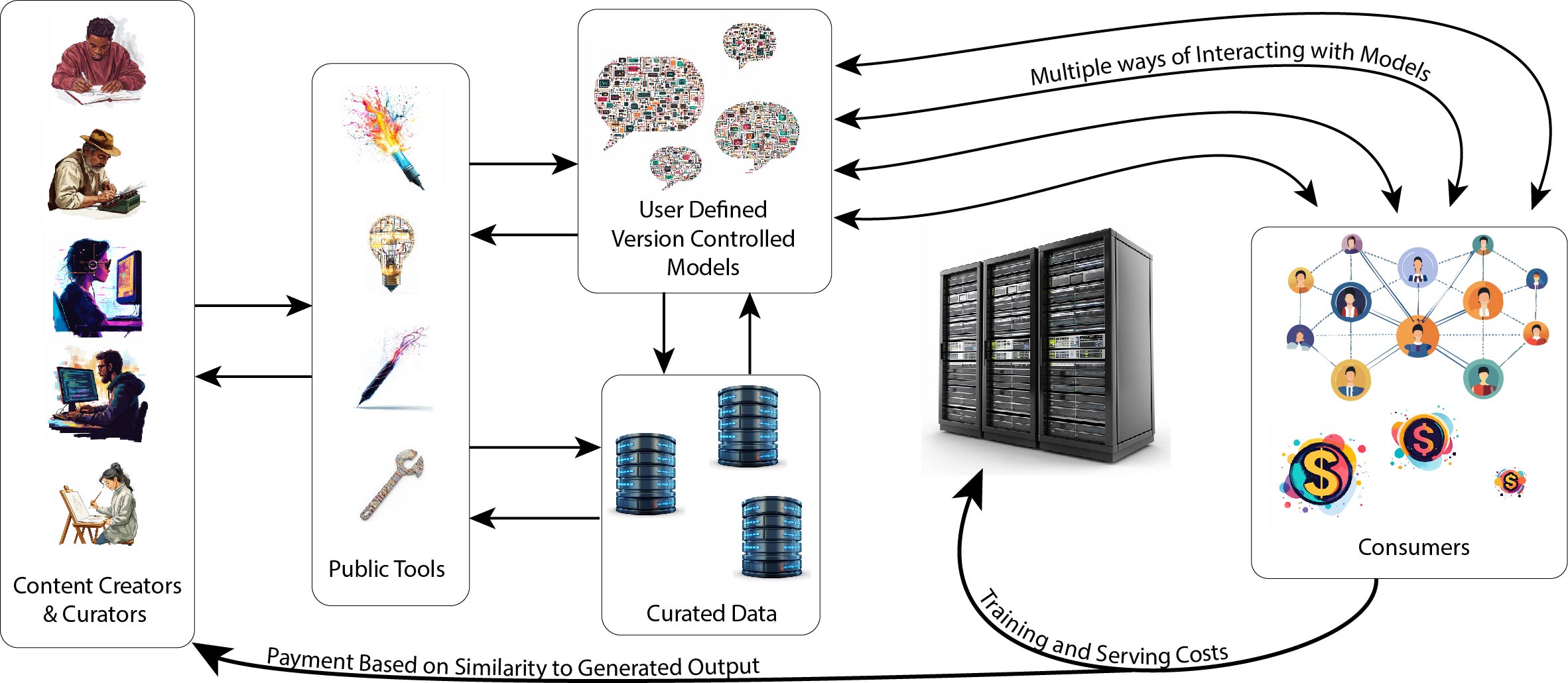}}
    \caption{Egalitarian AI Environment Concept}
    \label{fig:egalitarian}
\end{figure}

Figure~\ref{fig:egalitarian} shows an overview of the egalitarian environment as it might exist. In this ecosystem, content creators and curators actively contribute tagged data to open training databases, ensuring high quality and transparency. This open model allows the development of diverse tools that interact with the data in unforeseen ways, fostering a dynamic and responsive environment.  As user needs evolve, new tools can be seamlessly integrated into the existing open-source infrastructure. This flexibility extends to  novel user interfaces, potentially moving beyond the prompt-based interactions commonly associated with current generative AI models. The ecosystem is supported by voluntary contributions, which fund the infrastructure and provide revenue for content creators.  Finally, all models, including superseded versions, remain accessible and downloadable, ensuring continuity and preventing disruptions.

This model should also emphasize the importance of community governance and ethical considerations. The Wikimedia Foundation's commitment to neutrality, verifiability, and the free flow of information could serve as guiding principles for the development and deployment of egalitarian generative AI. The establishment of clear guidelines and mechanisms for addressing issues such as misinformation, bias, and harmful content would be crucial to ensure that these models are aligned with societal values and promote the public good. By drawing inspiration from the success of Wikipedia and the Wikimedia Foundation, we can envision a future where Generative AI serves as a tool for collective intelligence and the democratization of knowledge, rather than a means for profit and control. 
\section{Challenges}

The development, training, and deployment of open-source foundational generative AI models using egalitarian principles presents a distinct set of challenges. Large amounts of data must be ethically collected, and models have to be trained and deployed in a way that reflects community values.

\subsection{Data}
\label{sec:data-challenge}
Assembling a dataset of sufficient size and quality~\cite{gpt3_burruss_2020} as those used by OpenAI, Google, Anthropic, and Mistral may be impossible, since an egalitarian dataset would not use copyrighted material without permission. It may require novel approaches to data collection, curation, and licensing, potentially involving partnerships with public institutions, academic bodies, and open-source communities.

However, a reliance on crowd-sourced data and volunteer moderators introduces the question of oversight. Who are the gatekeepers responsible for ensuring the quality of data used to train these egalitarian models?  While a distributed, community-driven approach is more democratic and open than private corporations, such an approach may have problems with potential biases and inconsistencies in data curation.  All individuals carry their own perspectives and interpretations, which would inevitably influence the data and any subsequent model trained on the data.

An additional question to ask is: is it even feasible for volunteers to effectively curate such datasets?  The task would almost certainly require the development of automated tools to assist human curators in identifying, tagging, and de-duplicating content which could in turn introduce biases and result in significant model performance limitations.

As these are foundation AI models that can be used in many different downstream applications, curating content without a clear understanding of its intended use introduces additional complexity.  The appropriateness of certain data can vary wildly depending on the specific application context.  For instance, a dataset containing tax law and monetary policy might be valuable for training a model focused on retirement planning, but perhaps inappropriate for a model designed to generate children's stories.  This disconnect between data curation and downstream tasks will require flexibility and nuance in any kind of useful data curation and  annotation process.

This challenge is further compounded by the evolving nature of language and cultural norms. What might be considered acceptable language in one context or at one point in time could easily become offensive or outdated later.  Therefore, ongoing data monitoring and re-evaluation will be essential to ensure continued applicability. This adds another layer of complexity to the already demanding task of curating massive datasets for egalitarian generative AI models.

\subsection{Models}
\label{sec:model-challenge}

Training large Generative AI models requires substantial computational resources including specialized hardware. The now-obsolete GPT3 Davinci model is estimated to have cost \$4.6 million and required approximately 355 GPU-years to train~\cite{enwiki-gpt3-2024}. Such compute is typically the domain of well-funded commercial entities or governments.  Egalitarian approaches might involve offloading compute to idling computers and servers, a modern version of project SETI@home~\cite{korpela2001seti}, or by partnering with organizations that have access to high-performance computing clusters.  Research into optimization techniques and novel distributed training methods could also help.

Given the emphasis on ethical considerations and potential biases, a broader range of model evaluation criteria may be needed. This could involve developing new methodologies and benchmark datasets that assess fairness, transparency, and social impact, alongside traditional measures of accuracy and fluency.  A diverse and skilled volunteer community would provide a distinct advantage in this type of process, since the volunteer pool would include a greater variety of perspectives to these tasks.

Such limitations on data could potentially encourage a shift towards quality over quantity.  Smaller but high-quality curated datasets could be focused on extracting maximum value from well-tagged records.  This could lead to the development of specialized models, each trained on a specific subset of the data, resulting in a unique approach to the \enquote{mixture of experts} that collectively covers a wide range of knowledge and capabilities~\cite{cai2024survey}.  Such an approach could offer unique advantages, allowing for greater control over model behavior and potentially exceeding the performance of general-purpose models in specific domains.

Deployment strategies for egalitarian models open up interesting possibilities.  While the OpenAI model of providing API access is certainly an option, inference might be distributed among multiple computers in ways that might be similar to the distributed training discussed above~\cite{borzunov2024distributed}. Open source models can also be downloaded, using the Hugging Face model. This would empower users to have greater control over their AI interactions and potentially foster further innovation within the community.

Recent history has shown that open-source and even content protected by copyright has been exploited to provide training material for for-profit entities.  Strong community governance and well-crafted licensing agreements could help ensure that the egalitarian principles underpinning the project are upheld and that the models are used in a way that benefits society as a whole.

\subsection{Learning from Community-Driven Initiatives}

The challenges faced in creating egalitarian generative AI models can find inspiration and guidance in the successes of existing community-driven projects.  Wikimedia, with its collaborative Wikipedia and Wikidata models, demonstrates the power of harnessing diverse perspectives for knowledge creation and curation. Hugging Face, an open-source platform for sharing and collaborating on machine learning models, highlights the potential of community-driven development and deployment.  These examples, along with other community-focused efforts, offer valuable lessons in addressing the key issues outlined earlier.

Wikimedia's success lies in its established systems for organizing and curating vast amounts of information.  Adapting these strategies, including clear guidelines, version control, and dispute resolution mechanisms, can be a model in data curation for Generative AI models.  Wikimedia's experience in coordinating volunteers, motivating contributions, and fostering a sense of ownership would be critical in building a strong and engaged community around an egalitarian AI project.

Hugging Face's platform not only provides access to pre-trained models but also facilitates collaborative training and deployment.  This approach, called DeDLOC, distributes the computational load of training and inference across ad-hoc networks of computers. The effectiveness of this approach was  demonstrated by the creation of \textit{sahajBERT}, a high-performing Bengali language model trained by 40 volunteers~\cite{diskin2021distributeddeeplearningopen}.

A community approach to training data is also emerging. In March of 2024, the French start-up Pleias created a 500 billion-word training corpus composed on non-copyrighted material.\footnote{\url{https://www.euronews.com/next/2024/04/02/this-french-start-up-just-proved-openai-wrong-it-claims-you-can-train-ai-on-non-copyrighte}} This corpus is available on the Hugging Face Platform.\footnote{\url{https://huggingface.co/blog/Pclanglais/common-corpus}}

An egalitarian generative AI ecosystem, drawing inspiration from the success of Hugging Face, Wikipedia and other community-driven entities, could provide a way to address the challenges described in Sections \ref{sec:data-challenge} and \ref{sec:model-challenge}. Egalitarian approaches are already becoming a potential alternative to the current for-profit paradigm, and could lead to a more inclusive, transparent, and equitable AI landscape.

Ultimately, the success of egalitarian generative AI hinges on fostering a sustainable ecosystem that values transparency, accountability, and fair compensation for content creators. Establishing clear agreements with downstream providers that use these models with a realistic framework to remunerate content providers who provided the training data and the technical contributors who built and deployed the models. An egalitarian approach provides an environment where researchers, developers, and the public alike can scrutinize, improve, and contribute to the responsible development of AI technologies. This approach not only ensures a more equitable distribution of value but also paves the way for accelerated innovation and a future where AI truly benefits society as a whole.

\section{Conclusion} 







In a 2022 article for IEEE Intelligent Systems, Michael Wooldridge, Professor of Computer Science at the University of Oxford wrote:

\begin{displayquote}
    \textit{\enquote{The scale of Big AI systems -- the datasets and compute resources required to train them, and the salaries paid to the researchers that design and build them mean that \textbf{Big AI is effectively owned by Big Tech}. For the most part, individual universities and research groups simply cannot compete, even if the source code is released, this means relatively little without the data and compute resources to train the model. \textbf{Should we be concerned?  I think yes.} We would surely be concerned if physics or chemistry research could only be done by private companies. For all sorts of reasons, I believe we should be very concerned that \textbf{a key future technology is proprietary and closed—and that we and our students can only really access it if we are prepared to go behind locked doors}.}~\cite{wooldridge2022welcome}}
\end{displayquote}


The internet, as it stands, is largely a reflection of the dominant culture -- white, male, and increasingly corporate. Commercial generative AI models, trained on  this internet data inherit and perpetuate these biases. The companies that produce these models are primarily driven by returning the investment of their shareholders. They are not incentivized to do more than the minimum to acknowledge and include the voices and experiences of underrepresented groups.

As Wooldridge argues, the concentration of AI development within Big Tech creates an environment where key future technologies are proprietary and closed. This limits innovation and reinforces existing power structures. The development of a more inclusive and fair technological landscape requires a shift away from corporate dominance, towards systems that can grow and evolve according to the needs of users, not just the owners.

For technology to be more inclusive and fair, it has to have the ability to grow according to user's needs, not the owners. An example of this was the emergent use of  the hashtag (\#) by the users of Twitter, and its subsequent incorporation into the codebase. The hashtag allowed for \enquote{counterpublics} such as Black Twitter and Occupy Wall Street to emerge spontaneously~\cite{gutenberg2024jarvis, graham2016content}. These and other online communities that represent minority and marginalized voices are rarely a product of intentional design. Rather, they emerge within the affordances of the technology these individuals and groups are able to adapt to their needs.

As long as generative AI remains primarily within the domain of large, corporate interests, it is likely to perpetuating the biases of the dominant culture. To achieve truly representative and inclusive AI, we need to foster the development of egalitarian technological ecosystems that are structured to support diverse perspectives. This means supporting open-source initiatives, promoting data sovereignty, and advocating for policies that prioritize ethical considerations and user rights over corporate profit.

While the vision laid out in this paper may not represent the definitive form of an egalitarian AI ecosystem, we believe that it provides an initial roadmap for navigating the complexities of creating a more equitable and user-centric approach to AI development. By embracing the principles of collaboration, community ownership, and mutual benefit, we can build a future where AI serves as a tool for collective empowerment and creative expression for the many, rather than a means for the extraction of value by the few. Like the FOSS movement before it, such an endeavor could allow generative AI to more prioritize the needs and interests of users. 

It is becoming clear that the current trajectory of generative AI may be becoming unsustainable. We call for a radical re-imagining of the relationship between AI and society, one where individuals are not merely passive consumers but can be active participants.

\end{document}